\documentstyle[11pt]{article} 
\newcommand{\bmk}{\mbox{\boldmath $k$}}

\begin{document} 

\begin{center} 

\hspace*{10mm} 

{\bf Relativistic Model for two-band Superconductivity}\\

\vspace{.5cm} 

{Tadafumi Ohsaku}\\ 

\vspace{.3cm} 
{\em Research Center for Nuclear Physics, Osaka University, Osaka, Japan}\\

\end{center} 
\vspace{0.5cm} 

Recent investigation of the superconductivity in ${\rm MgB}_{2}$
gived a large influence upon condensed matter physics~[1].
One of the characteristic aspects of the superconductivity is 
rather high critical temperature $T_{c}$ (39K) in electron-phonon mechanism.
Therefore, people speculate that, there is a mechanism
which enhances the superconductivity and $T_{c}$.
After the experimental discovery of the superconductivity in ${\rm MgB}_{2}$,
various theoretical studies were performed.
Several two-band theories were also dedicated to explain the superconductivity.
The two-band models of superconductivity have the origin in 
the papers of Suhl, Matthias and Walker~[2], and Kondo~[3], 
and they were applied to various systems under various situations. 
Quite recently, theoretical studies of two-band superconductivity 
in ${\rm MgB_{2}}$ appeared~[4], 
and the experimental evidence for two-band superconductivity 
in ${\rm MgB_{2}}$ was also obtained~[5]. 
However, there is no quantitative understanding to the effect of 
the lower band in two-band superconductivity.
In this paper, by using the relativistic fermion model,
we clearize the effect of the lower band in the superconductivity.

${\rm MgB}_{2}$ has honeycomb lattice structure similar to graphite.
As a consequence of its symmetry, the band structure of it has two
degeneracy points per a first Brillouin zone ( two conical intersections
between the upper band and lower band ). Then the relativistic dispersion arises
in the vicinity of these degeneracy points~[6]. 
We employ the Dirac fermion model with a BCS-like interaction
to study the superconductivity~[7]. We start from the following Lagrangian:
\begin{eqnarray}
{\cal L}(x) &=& \bar{\psi}(x)i\gamma^{\mu}\partial_{\mu}\psi(x)+\frac{G_{0}}{2}(\bar{\psi}(x)\psi(x))^{2}.
\end{eqnarray}
After usual Gor'kov factorization was performed to (1), 
we yield the (2+1)-dim. relativistic Gor'kov equation
(in the finite-temperature Matsubara formalism):
\[
\left(
\begin{array}{cccc}
-\gamma^{0}(\partial_{\tau}-\mu)+i\gamma^{k}\partial_{k} & \Delta(x) \\
\bar{\Delta}(x) & -(\gamma^{0})^{T}(\partial_{\tau}+\mu)+i(\gamma^{k})^{T}\partial_{k}
\end{array}
\right)
\left(
\begin{array}{cccc}
{\cal S}(x,y)_{\alpha\beta} & -{\cal F}(x,y)_{\alpha\beta} \\
-{\bar {\cal F}}(x,y)_{\alpha\beta} & -{\cal S}(y,x)_{\beta\alpha}
\end{array}
\right)
\]
\begin{equation}
=
\left(
\begin{array}{cccc}
\delta^{(3)}(x-y) & 0 \\
0 & \delta^{(3)}(x-y) 
\end{array}
\right).
\end{equation}
The chemical potential $\mu$ determines the position of the Fermi level measured from the degeneracy point.
We decompose the mean fields as follows:
\begin{eqnarray}
\Delta = (\Delta^{S}+\Delta^{V}_{\mu}\gamma^{\mu})C, \qquad \bar{\Delta} = -C^{-1}(\Delta^{S*}+\Delta^{V*}_{\mu}\gamma^{\mu}),
\end{eqnarray}
where $S$ indicates scalar, while $V$ indicates vector. $C$ is the charge conjugation matrix.
Hereafter, we concentrate on the case of the scalar pairing.
From the solution of the Gor'kov equation, we obtain the gap equation in the following form: 
\begin{eqnarray}
1 &=& \frac{G_{0}}{2}\int\frac{d^{2}\bmk}{(2\pi)^{2}}\Bigl(\frac{1}{2E_{+}}\tanh\frac{\beta}{2}E_{+}+\frac{1}{2E_{-}}\tanh\frac{\beta}{2}E_{-}\Bigr), \\
E_{\pm} &=& \sqrt{(|\bmk|\mp\mu)^{2}+|\Delta^{S}|^{2}}.
\end{eqnarray}
Here, the first term in the integrand is the contribution coming from the positive energy state (upper band),
while the second term is the contribution coming from the negative energy state (lower band).
To see the effect of the lower band, we also treat the gap equation 
in which the second term of integrand is dropped (single band case, the "no sea" approximation) in Eq. (4).

The results of the solutions of our gap equations are summarized as follows: 
(1)The solutions obtained from both the two-band and single-band cases always show the BCS behaviour:
The strength of the gap linearly depends on the cutoff $\Lambda$ 
and exponentially depends on the coupling $G_{0}$.
The solutions always fulfill the BCS universal relation $|\Delta(T=0)|/T_{c}=1.76$.
(2)The ratio $T^{two-band}_{c}/T^{single-band}$ becomes 1.57. 
Therefore, the lower band effect gives the sizable enhancement in the supercoductivity.
In fact, these results are expected from previous studies of relativistic superconductivity
we have published~[7].
If we take $\mu=1$eV (which is determined by band width) , $T^{two-band}_{c}$ becomes 38K 
while $T^{single-band}_{c}$ becomes 24K.
Upperbound of $T_{c}$ in electron-phonon mechanism is estimated by McMillan
as 20K$\sim$30K~[8]. Our result gives a support to the two-band BCS mechanism in ${\rm MgB}_{2}$.

We also have studied the possibility arising from a repulsive interaction
between particles ( the Kohn-Luttinger effect~[9] ) in the same model, 
by employing the method of the Bethe-Salpeter theory.
Summary of our results is given in table.
Detailed discussion of this work will be published elsewhere~[10].

\begin{table}
\caption{List of solutions of various pairings.}
\begin{tabular}{ccc}
pairing symmetry & Gor'kov-BCS ( $G_{0}>0$ ) & Bethe-Salpeter-RPA ( $G_{0}<0$ ) \\
\hline
scalar & possible & no solution \\
0th-component of vector & no solution & no solution \\
spatial-component of vector & no solution & no solution \\
\end{tabular}
\end{table}

\end{document}